\documentclass[12pt]{article}
\pdfoutput=1
\usepackage{cite}
\usepackage{amssymb}
\usepackage{amsmath}
\usepackage{bm} 
\usepackage{graphicx}
\usepackage{color}
\usepackage{xcolor}
\usepackage{dsfont}
\usepackage{enumerate}
\usepackage{hyperref}
\usepackage{setspace}
\usepackage[latin1]{inputenc}
\definecolor{nicered}{rgb}{0.7,0.1,0.1}
\definecolor{nicegreen}{rgb}{0.1,0.5,0.1}
\hypersetup{colorlinks,citecolor= nicegreen,linkcolor= nicered}

\newcommand{\be}  {\begin{equation}}
\newcommand{\ee}  {\end{equation}}

\def\e6{E(6)}
\def\10{SO(10)}
\def\21{SA(2) $\otimes$ U(1) }
\def\321{$\mathrm{SU(3) \otimes SU(2) \otimes U(1)}$ }

\def\422{SA(4) $\otimes$ SA(2) $\otimes$ SA(2)}
\def\roughly#1{\mathrel{\raise.3ex\hbox{$#1$\kern-.75em
      \lower1ex\hbox{$\sim$}}}} \def\lsim{\roughly<}
\def\gsim{\roughly>}

\def\lsim{\raise0.3ex\hbox{$\;<$\kern-0.75em\raise-1.1ex\hbox{$\sim\;$}}}
\def\gsim{\raise0.3ex\hbox{$\;>$\kern-0.75em\raise-1.1ex\hbox{$\sim\;$}}}
\nopagebreak
\allowdisplaybreaks
\begin{document}
\begin{titlepage}


  \newcommand{\AddrIPM}{{\sl \small School of physics, Institute for
      Research in Fundamental Sciences (IPM),\\ \sl \small
      P.O. Box 19395-5531, Tehran, Iran}}
  \vspace*{0.5cm}
\begin{center}
  \textbf{\large A model for large non-standard interactions of neutrinos leading to the LMA-Dark solution  }

  Yasaman Farzan\footnote{e-mail address:{\tt yasaman@theory.ipm.ac.ir}}
  \vspace*{0.4cm}\\
  \AddrIPM.
  \vspace{1cm}\\
\end{center}
\vspace*{0.2cm}
\begin{abstract}
 \onehalfspacing
It is well-known that in addition to the standard LMA solution to solar anomaly, there is another solution called LMA-Dark  which requires Non-Standard Interactions  (NSI) with effective couplings as large as the Fermi coupling. Although this solution satisfies all the bounds from various neutrino oscillation observations and even provides a better fit to low energy solar neutrino spectrum, it is not as popular as the LMA solution mainly because no  model compatible with the existing bounds has been so far constructed to give rise to this solution. We introduce a model that provides a foundation for such large NSI with strength and flavor structure required for the LMA-Dark solution. This model is based on a new $U(1)^\prime$ gauge interaction with a gauge boson of mass $\sim 10$ MeV under which quarks as well as the second and third generations of leptons are charged. We show that observable effects can appear in the spectrum of supernova  and high energy cosmic neutrinos. Our model predicts a new contribution to the  muon magnetic dipole moment and new rare meson decay modes.

\end{abstract}
\end{titlepage}
\setcounter{footnote}{0}
\section{Introduction
}\label{sec:intro}
Three neutrino mass and mixing scheme has been established as the standard solution to the solar and atmospheric neutrino anomalies.  In general, the implicit assumption is that
the alternative solutions to these anomalies ({\it e.g.,} the magnetic transition moment of neutrinos) are now ruled out   and can at most  provide  only a subdominant effect  yet  to be resolved. However, the so-called LMA-Dark solution, which is based on  relatively large non-standard neutral current  interactions of neutrinos with matter,  is  an exception that defies this general assumption.

Surprisingly, the LMA-Dark solution not only passes all the bounds from various neutrino oscillation experiments but  is also  considered as  one of the solutions for the suppression of the upturn on low energy spectrum of solar neutrinos  which is in a mild tension  with the standard neutrino oscillation scenario \cite{upturn}.
In other words, the compatibility of LMA-Dark solution with the neutrino data is even slightly better than the standard three neutrino oscillation scheme without Non-Standard Interaction (NSI). The reason for this solution staying in the shadow of the LMA solution is mostly  theoretical. In fact, there is no shortage of beyond standard models that give rise to NSI of neutrinos with matter \cite{Tommy,Miranda}. The ratio of new effects within beyond  Standard Model (SM) scenarios on forward scattering amplitude of neutrinos on matter to the SM contribution is expected to be given by $(g_X^2 m_X^{-2})/ G_F$, where $g_X$ and $m_X$ are respectively the coupling and mass of the new particle whose exchange leads to an effective four-Fermi interaction between neutrinos and matter field. Taking $m_X\gg m_W$ to avoid bounds from direct  collider searches  of new particles  and requiring $g_X$ to be relatively small to remain in the perturbative region, we find that the  NSI effects are suppressed relative to the standard model effects.
For the  LMA-Dark solution, the NSI effects should be as large as the standard ones. Thus, in the framework of beyond standard models with heavy  particles ($m_X\gg m_W$), there is no theoretical basis for the LMA-Dark solution.
The purpose of this letter is to build a model or a class of models that provide such a basis.

 The model presented here is based on  a new $U(1)$ gauge interaction with a gauge boson of mass $m_{Z^\prime}\sim {\rm few}~10~{\rm MeV}$ and gauge coupling of $g_{Z^\prime}\sim 10^{-5}$ which couples to the second and third generations of leptons   (but not to the electron) as well as to the quarks. This gauge interaction leads to effective four-Fermi interactions with form and coupling strength required for the LMA-Dark solution.
We expect new observable effects  on the muon magnetic dipole moment, supernova evolution, trident neutrino production, meson decays and the
spectrum of high energy cosmic neutrinos \cite{Kamada:2015era,Cherry}. As shown in \cite{russians}, such a model can be tested by studying $ \mu+A\to \mu +A + Z^\prime$ where $A$ is a nucleus.

The letter is organized as follows: In sec. 2, the LMA-Dark solution is briefly reviewed. In sec. 3, the model is presented. In sec. 4, the potential signals of the model and observational bounds
are reviewed. A summary of results is presented in sec. 5.

\section{The LMA-Dark solution}
Within the standard model, the neutral current interactions are flavor diagonal and universal. Going beyond the standard model, the neutral current interactions can have a more general form. Integrating out  the heavy states, the correction to the four-Fermi neutral current interaction of  the neutrino pair with matter can be described by the following effective operator \be \label{NSI}
\mathcal{L}_{NSI}=-2\sqrt{2}G_F\epsilon_{\alpha \beta}^{f P}
(\bar{\nu}_\alpha \gamma^\mu L\nu_\beta)(\bar{f}\gamma_\mu P ~f) \ee
where $f$ is the matter field ($u, \ d$ or $e$), $P$ is the
chirality projection matrix and $\epsilon_{\alpha \beta}^{f P}$ is a
dimensionless matrix describing the deviation from the standard
model. For propagation of neutrinos in matter, only forward scattering is relevant so integrating out the intermediate states in the $t$-channel amplitude and using the effective four-Fermi interaction is perfectly justified even if the intermediate states are relatively light. Since the atoms in the medium in which neutrinos propagate (the Sun, the Earth and etc.) are non-relativistic, mainly the vector part of the interaction contributes to the effective potential of neutrinos in matter which can be described by the following combination:
$$ \epsilon_{\alpha \beta}^{f }\equiv \epsilon_{\alpha \beta}^{f L}+\epsilon_{\alpha \beta}^{f R}.$$
The  non-standard interactions have been invoked in \cite{upturn} to explain the suppression of upturn in low energy  solar neutrino spectrum. This new solution which is called the LMA-Dark solution requires a value for $\theta_{12}$ larger than $\pi/4$ and
\be -1.40<\epsilon_{ee}^u-\epsilon_{\mu \mu}^u<-0.68 \ \ {\rm and} \ \ -1.44<\epsilon_{ee}^d-\epsilon_{\mu \mu}^d<-0.87 \ \ {\rm at~3\sigma ~C.L.} \label{range}\ee
 The  off-diagonal elements of $\epsilon_{\alpha \beta}$  and $\epsilon_{\mu \mu}-\epsilon_{\tau \tau}$ in the LMA-Dark solution can vanish. In \cite{Juno}, we have shown that the intermediate baseline reactor experiments such as JUNO can test this solution by determining  which octant  $\theta_{12}$ belongs to\footnote{Notice that the reactor experiments are not directly sensitive to neutral current NSI but they can probe charged current NSI \cite{ccNSI}. LMA-Dark solution does not however involve charged current NSI.}. NSI effects in long baseline experiments have been explored in \cite{shoemaker}.
Notice that the diagonal couplings of NSI at the LMA-Dark solution are as large as the standard ones. It is remarkable that such large values are still compatible with all the neutrino oscillation data \cite{Maltoni}. However, as shown in \cite{bounds},  the results of  the CHARM and NuTeV experiments rule out a significant (but not all) part of the range shown in Eq. (\ref{range}).
Notice however that to extract these bounds, it was implicitly assumed that the scale of new physics behind the effective couplings in Eq. (\ref{NSI})
was much higher than the typical energy exchange in these scattering experiments rendering the effective Lagrangian formalism  valid. If, as in the model presented in the next section, the masses
of new particles are below typical energies at scattering experiments, these bounds do not apply.
\section{The model}
In this section, we show that it is possible to build a class of models that give rise to NSI of form required for the LMA-Dark solution ($\epsilon_{\mu \mu}-\epsilon_{ee}\sim\epsilon_{\tau \tau}-\epsilon_{ee}\sim 1$) avoiding all the present bounds. The models are based on a spontaneously broken $U(1)$ gauge theory  with gauge boson $Z_\mu^\prime$ of mass of $O(10)$ MeV that couples to the second and third generations of leptons as well as the first generation of quarks. We denote the new gauge symmetry by $U(1)^\prime$. To avoid the  strong bounds on the coupling to the first generation of leptons, we assume that the $U(1)^\prime$ charges of the first generation of leptons vanish. Although  coupling to the charged leptons or the second  and third generations of quarks is not required for the LMA-Dark solution, we however need coupling to these fermions to cancel anomalies.  In summary, the following gauge couplings are the basis of models that lead to NSI required for the LMA-Dark solution
\be -g^\prime\left( Y_{L}\sum_{\alpha \in \{ \mu, \tau \}}  \bar{L}_{\alpha} \gamma^\mu L_{\alpha}+Y_{Q_1} \bar{Q}_1\gamma^\mu Q_1+Y_{u_1} \bar{u}_R\gamma^\mu u_R+ Y_{d_1}\bar{d}_R\gamma^\mu d_R \right) Z^\prime_\mu \in \mathcal{L} \label{necessary-couplings} \ee
where $L_\alpha$ and $Q_1$ are the doublets of leptons of flavor $\alpha$ and quarks of first generation, respectively.
Effects of couplings in Eq. (\ref{necessary-couplings})  on the forward scattering (or any scattering with $q^2 \ll m_{Z^\prime}^2$) can be described by the following effective four-Fermi interactions:
\be \mathcal{L}_{NSI}=-\frac{g^{\prime 2} Y_L}{m_{Z^\prime}^2} (\sum_{\alpha\in \{ \mu ,\tau \}} \bar{L}_\alpha \gamma^\mu L_\alpha )\left( Y_{Q_1} \bar{Q}_1 \gamma_\mu Q_1+Y_{u_1} \bar{u}_R\gamma_\mu u_R+Y_{d_1} \bar{d}_R\gamma_\mu d_R \right) \ . \label{four}\ee
Notice that we have taken the same couplings for $\mu$ and $\tau$. 
The mixing between $\nu_e$ and $\nu_\mu ~(\nu_\tau)$
in the neutrino mass matrix is not compatible with the $U^\prime (1)$ symmetry under which $\nu_\mu$ and $\nu_\tau$ are charged but $\nu_e$ is neutral.  The mixing of $\nu_e$ with $\nu_\mu$ and $\nu_\tau$ can be
generated only after spontaneous breaking of the   $U^\prime (1)$ gauge symmetry which requires  new scalar(s). We will return to this point later.

Equating $\mathcal{L}_{NSI}$ in Eqs. (\ref{NSI}) and (\ref{four}), we find
\be \epsilon_{\tau \tau}^u=\epsilon_{\mu \mu}^u=\frac{g^{\prime 2}}{m_{Z^\prime}^2} \frac{Y_L (Y_{Q_1}+Y_{u_1})}{2\sqrt{2} G_F} \label{EtauU}\ee
and \be \epsilon_{\tau \tau}^d=\epsilon_{\mu \mu}^d=\frac{g^{\prime 2}}{m_{Z^\prime}^2} \frac{Y_L (Y_{Q_1}+Y_{d_1})}{2\sqrt{2} G_F} \label{EtauD}.\ee
Requiring $\epsilon_{\mu \mu}^{u,d}\sim 1$, we find \be \label{Gprime} g^\prime \sim 7 \times 10^{-5} \frac{m_{Z^\prime} }{10~{\rm MeV}}.\ee
In order to avoid the bounds from big bang nucleosynthesis, the mass of $m_{Z^\prime}$ should be of order of 10 MeV or larger \cite{Kamada:2015era}.
Taking $\epsilon_{\mu \mu}^{u,d}\sim 1 $, we find that at neutrino scattering experiments with four-momentum transfer of $q$, the contribution of new gauge interaction to scattering amplitude relative to the standard amplitude is given by
$m_{Z^\prime}^2/(q^2-m_{Z^\prime}^2)$. Notice that we have used the signature $\eta^{\mu \nu}={\rm Diag}(1,-1,-1,-1)$ so the $t$-channel propagator is given by $1/(q^2-m_{Z^\prime}^2)$. To avoid significant deviation from the SM prediction at scattering experiments such as NuTeV, NOMAD and etc., $m_{Z^\prime}$ should be much smaller than the typical energy of these experiments which is of order of GeV
$$ 10 ~{\rm MeV} \stackrel{<}{\sim} m_{Z^\prime}\ll 1~{\rm GeV}.$$  Similar consideration holds valid for scattering of the muon beam (or hypothetical tau beam) on quarks.

Up to now, we have only studied the couplings that give rise to NSI relevant for neutrino propagation. Let us now discuss the $U(1)^\prime$ gauge couplings of all SM fermions and the  possibility of
the anomaly cancelation. Notice that $Y_L$,  $(Y_{Q_1}+Y_{u_1})$ and $(Y_{Q_1}+Y_{d_1})$ should be all of  the same sign to obtain positive value for effective couplings $\epsilon_{\mu \mu}^{u,d}$ and
$\epsilon_{\tau \tau}^{u,d}$ defined in Eqs (\ref{EtauU}) and (\ref{EtauD}) as  required for the LMA-Dark solution.
As is well-known in the presence of right-handed neutrinos, the anomalies of $\mathcal{L}_e$, $\mathcal{L}_\mu$, $\mathcal{L}_\tau$ and $\mathcal{B}/3$ (where $\mathcal{L}_\alpha$ is the lepton number of flavor $\alpha$ and $\mathcal{B}$ is the Baryon number) are equal.
  Thus, the combinations such as $\mathcal{B}-\mathcal{L}$, $\mathcal{L}_\mu-\mathcal{L}_\tau$ or $\mathcal{L}_\mu+\mathcal{L}_\tau-(2/3)\mathcal{B}$ are anomaly free  and can be gauged. In fact, there is rich literature on the $\mathcal{B}-\mathcal{L}$ and $\mathcal{L}_\mu-\mathcal{L}_\tau$ gauge theories. However,
these models do not reproduce the flavor structure of $\epsilon_{\alpha \beta}^u$ and $\epsilon_{\alpha \beta}^d$ matrix as required by the LMA-Dark solution because while  $\mathcal{B}-\mathcal{L}$ predicts opposite sign for the $U(1)^\prime$ charges of leptons and quarks, $\mathcal{L}_\mu-\mathcal{L}_\tau$ predicts null coupling to quarks.
Notice that the LMA-Dark solution differentiates between first generation of the leptons and the rest, $\epsilon_{\mu \mu}^{u,d}\simeq \epsilon_{\tau \tau}^{u,d}
\neq \epsilon_{ee}^{u,d}$ so we should assign different $U(1)^\prime$ charges to these leptons. Similarly to the lepton sector, let us define $\mathcal{B}_i$ as quark number of generation $i$. Combinations of form
\be \label{combination}\mathcal{L}_\mu+\mathcal{L}_\tau+\mathcal{B}_1- a\mathcal{B}_2-(3-a)\mathcal{B}_3 \ee (with an arbitrary $a$)\footnote{More general combination of
 $\mathcal{L}_\mu+c\mathcal{L}_\tau+b\mathcal{B}_1- a\mathcal{B}_2-(1+c+b-a)\mathcal{B}_3$ can also serve this purpose as long as arbitrary real numbers $c$ and $b$ satisfy
 $c\simeq 1$ and $b>0$. We however restrict ourselves to the less general combination shown in Eq. (\ref{combination}) to avoid unnecessary complexity. The case of $c=-1$ and $b=-a$ has been proposed and explored in \cite{Heeck} to explain the LHC flavor anomalies. } are also anomaly free. Gauging such a  symmetry leads to
\be \ Y_{Q_1}=Y_{u_1}=Y_{d_1}=1/3 \ , \ \ Y_{Q_2}=Y_{u_2}=Y_{d_2}=-a/3 \  \ {\rm and}  \ \  Y_{Q_3}=Y_{u_3}=Y_{d_3}=-1+a/3 \label{quarks}\ee and \be Y_e=Y_{L_e}=0 \  \  {\rm and} \ \   Y_\mu=Y_\tau=Y_{L_\mu}=Y_{L_\tau}=1 \ , \label{leptons}\ee
where $Y_e$, $Y_\mu$ and $Y_\tau$ are $U(1)^\prime$ charges of the corresponding right-handed leptons.
Notice that with this assignment, the left-handed and right-handed fermions of each generation have the same $U(1)^\prime$ charge so the diagonal elements of mass matrices do not break the
$U(1)^\prime$ symmetry but the off-diagonal elements of the mass matrices (except the $\tau \mu$ element of neutrino mass matrix) break $U(1)^\prime$. As a result, to obtain mixing between different generations, new scalar(s) with proper $U(1)^\prime$ charges should be introduced to couple to the fermions. The off-diagonal elements can be obtained after spontaneous breaking of electroweak and  $U(1)^\prime$ symmetries.

Let us first discuss neutrino masses. The Dirac neutrino masses can be obtained from the following Yukawa couplings
\be \lambda_1 \bar{N}_1H^T c L_e+ \lambda_2 \bar{N}_2H^T c  L_\mu+ \lambda_3 \bar{N}_3H^T c  L_\tau+ \lambda_4 \bar{N}_2 H^T c L_\tau+ \lambda_5 \bar{N}_3H^T c  L_\mu+{\rm H.c.}\ee
where $N_1$ like $L_e$ is neutral under $U(1)^\prime$ but $N_2$ and $N_3$ are charged: $Y_{N_2}=Y_{N_3}=Y_{L_\mu}=Y_{L_\tau}=1$. $c$ is two by two antisymmetric matrix  with $c_{12}=-c_{21}=1$. Notice that by changing basis, either $\lambda_4$ or $\lambda_5$ can be set to zero but in general one of them remains nonzero and  mixes the $\mu$ and $\tau$  flavors
with each other. However, to mix $\nu_\mu$ and $\nu_\tau$ with $\nu_e$, we need extra scalars to break $U(1)^\prime$. If we take only Dirac mass term, we shall encounter two problems:
1) Smallness of neutrino masses cannot be explained. 2) For Dirac neutrinos, we expect $m_{N_i}\sim m_\nu$. Considering that $N_2$ and $N_3$ are charged under $U(1)^\prime$ they can be produced  via
$U(1)^\prime$ gauge interaction in the early universe and contribute to the relativistic degrees of freedom $(\Delta N_{eff})$ on which there are strong bounds.
To solve both problems, we can invoke the seesaw mechanism. Majorana mass terms for $N_2$ and $N_3$ require a new scalar ($S_1$) with  $U(1)^\prime$  charge  $Y_{S_1}=-2$. To mix $\nu_e$ with the rest we introduce another scalar $S_2$ with  a $U(1)^\prime$ charge of $Y_{S_2}=-1$. We can then have the following terms in the  potential
$$M_1 N_1^TcN_1+S_1(A_2 N_2^TcN_2+A_3 N_3^TcN_3+A_{23} N_2^TcN_3)+S_2( B_2  N_1^TcN_2 +B_3 N_1^TcN_3)+{\rm H.c}.$$
where $A_i$ and $B_i$ are dimensionless couplings.
The mass of the $Z^\prime$ boson receives a contribution from the  Vacuum Expectation Value (VEV) of $S_1$ and $S_2$: $g^{\prime 2} (Y_{S_1}^2 \langle S_1 \rangle^2+Y_{S_2}^2 \langle S_2 \rangle^2)$.
However, since $S_1$ and $S_2$ are electroweak singlets, their VEVs cannot contribute to the masses of the $W$ and $Z$ gauge bosons.

Let us now discuss mixing in the quark sector considering $U(1)^\prime$ charges shown in Eq. (\ref{quarks}).
To minimize the number of required scalars, we can choose the value of $a$ such that the $U(1)^\prime$ charges of two generations of quarks become the same. It is 
straightforward to confirm that  taking $a=3/2$, $a=4$ or $a=-1$ serves this purpose.  Let us suppose that the $U(1)^\prime$ hypercharges of the $i$ and $j$
generations are the same but different from that of the $k$ generation. Going to the mass basis, the $ij$ element of the $Z^\prime$
coupling will be zero but the $ik$ and $jk$ elements will not vanish and can lead to flavor changing neutral currents. To avoid too large effects on the $K-\bar{K}$ and 
$D-\bar{D}$ mixing, we can set $a=-1$, rendering $U(1)^\prime$ charges of first and second generations the same. Thus, the Cabibbo mixing between first and 
second generations does not require breaking $U(1)^\prime$. To mix the third generation of quarks with the rest, 
we can introduce a second Higgs doublet $H^\prime$ with the same electroweak quantum numbers as those of the  SM Higgs but nonzero $U(1)^\prime$ charge given by $Y_{H^\prime}=Y_{Q_{1,2}}-Y_{d_3}=Y_{Q_{1,2}}-Y_{u_3}=5/3$.
With this new scalar, we can write Yukawa couplings of form $ H^{\prime\dagger}c\bar{Q}_3 u_{1,2}$ and $\bar{d}_3 H^{\prime \dagger} Q_{1,2}$. As a result, the VEV of $H^\prime$ will lead to  the $tc$, $tu$ as well as the $bd$ and $bs$ elements of the corresponding mass matrices but the $ct$, $ut$, $db$ and $sb$ elements of the same matrices vanish. The vanishing elements can be easily accommodated as we have the freedom to rotate the quarks with  arbitrary unitary matrices.  If the components of $H^\prime$  are not too heavy, they can be produced
at the LHC and decay to quark pairs of different generations. Moreover, they can lead to new tree level flavor changing low energy effects. If $H^\prime$ is heavy enough
($m_{H^\prime}\gg 100$~GeV), the bounds can be avoided but its VEV has to be smaller than $\langle H\rangle$ in the SM because, being electroweak doublet, it also contributes to the $Z$ and $W$ masses.
This can be achieved by introducing another scalar, $S$, with $U(1)^\prime$ charge of $-5/3$ and with the following potential
\be -m_S^2 |S|^2+\lambda_S |S|^4+m_{SHH^\prime} SH^\dagger H^\prime+m_{H^\prime}^2 H^{\prime \dagger}H^\prime,\ee
with $0<m_S^2\ll m_{H^\prime}^2$ and $m_{EW}^2<m_{H^\prime}^2$. We will then have
\be \langle S\rangle =\left(\frac{m_S^2}{2\lambda_S}\right)^{1/2} , \ \ \ \langle H^\prime \rangle =-m_{SHH^\prime} \frac{\langle H\rangle \langle S\rangle}{2 m_{H^\prime}^2}.
\ee  Taking $m_{SHH^\prime}, \langle H\rangle \ll m_{H^\prime}$ and $\langle S\rangle \stackrel {<}{\sim}\langle H\rangle$, we obtain  $\langle H^\prime \rangle \ll  m_{H^\prime},m_{EW}, \langle S\rangle$. Adding a quartic term for $H^\prime$ with a small coupling only slightly shifts $\langle H^\prime \rangle$. Notice that there is  a small mixing between $S$ and $H^{\prime 0}$ given by $m_{SHH^\prime} \langle H\rangle /m_{H^\prime}^2\ll 1$. A small mixing  also appears between $H^\prime$ and $H$ given by $m_{SHH^\prime} \langle S\rangle /m_{H^\prime}^2\ll 1$.    The mass of $Z^\prime$ boson will be given by
\be\label{Zmass} m_{Z^\prime}=g^\prime \left(Y_{S_1}^2 \langle S_1 \rangle^2+Y_{S_2}^2 \langle S_2 \rangle^2+Y_{S}^2 \langle S \rangle^2+Y_{H^\prime}^2 \langle H^\prime \rangle^2\right)^{1/2}.\ee
Taking $m_{Z^\prime}\sim 10 $ MeV and $g^\prime \sim 7 \times 10^{-5}$, we find that the VEV of electroweak  singlet scalars should be of order of few hundred GeV. In particular, $\langle S_1\rangle \sim \langle S_2\rangle \sim 100$ GeV leads to right-handed neutrino masses of order of 100 GeV. Although, the masses are within the LHC reach, the $g^\prime$ coupling is too small
to lead to a significant production of these particles at the LHC.  If the $H^\prime$ components are not too heavy, they can be produced at the Run II of the LHC by electroweak interactions and lead to two jets of different flavors.
\section{Observational bounds and potential signals of the model}
In this section, we review the possible observational signals within this model and  discuss the existing bounds.

{\it Meson decay:} Taking $Z^\prime$ lighter than the pion, Hadronic decay modes for $Z^\prime$ will be forbidden and  the  tree level decays into $\nu_\mu\bar{\nu}_\mu$ and
$\nu_\tau\bar{\nu}_\tau$ pairs will constitute the dominant decay mode.  In this range of parameters, $Z^\prime$ can be produced in the meson decays and show up as missing energy. Some examples are
$K^+\to \mu^+ +\nu_\mu +Z^\prime$, $K^+\to e^+ +\nu_e +Z^\prime$, $\pi^+\to \mu^+ +\nu_\mu +Z^\prime$ and $\pi^+\to e^+ +\nu_e +Z^\prime$. The bound that has been obtained from the meson decay  on the couplings of
such a particle is of order of $10^{-3}$ \cite{Lessa} which is more than one order of magnitudes above the value we consider here (see Eq. (\ref{Gprime})). Notice however that studying meson decays with higher accuracy can eventually provide a route to test
these models. Since the branching ratios  of these processes are proportional to $g^{\prime 2}$, to probe these effects precision on the branching ratios should be improved by a factor of $\sim 200$.

{\it Mixing between $Z^\prime$ and $\gamma$:} Considering that the charged leptons as well as the quarks couple both to the photon and the $Z^\prime$ boson,  a mixing will be obtained between ordinary photon and the $Z^\prime$ boson at one-loop level given
by $\epsilon \sim g^\prime e/(8 \pi^2)$. Such a mixing can affect the neutrino scattering on the electron. There are strong bounds from such interaction from Borexino experiment. The corresponding bound on the
$\mathcal{B}-\mathcal{L}$ models has been studied in \cite{Borexino}. Similarly to the $\mathcal{L}_\mu-\mathcal{L}_\tau$ model discussed in \cite{Kamada:2015era}, the bound  in our case has to be corrected by a factor of $(1/0.66)^{1/4}$ to account for the fact that
only the $\nu_\mu$ and $\nu_\tau$ components of the solar neutrino flux couple to $Z^\prime$. From the bound shown in Fig 8 of \cite{Borexino}, we obtain $(g^\prime e \epsilon)^{1/2}<10^{-5}$. Taking
$g^\prime\sim 7\times 10^{-5}$, we find $(g^\prime e \epsilon)^{1/2}\sim 2\times 10^{-6}$ which is well below this bound. The bound from  GEMMA \cite{Borexino} does not apply for our model because  $\nu_e$ does not couple to $Z^\prime$. The beam dump experiments, which provide strong bounds on the $\mathcal{B}-\mathcal{L}$ gauge theories are insensitive to this model because the produced $Z^\prime$ dominantly
decays into $\nu \bar{\nu}$ pairs which, unlike the electron positron pair expected from the $Z^\prime$ decay within the $\mathcal{B}-\mathcal{L}$ gauge models, leave no trace in detectors.
In the parameter range of interest with $m_{Z^\prime}\sim 10$~MeV, there is no significant bound on the coupling of neutrinos to quarks \cite{Escrihuela:2011cf}.

{\it Big bang nucleosynthesis:} Decay of $Z^\prime$ into $\nu_\mu \bar{\nu}_\mu$ and $\nu_\tau \bar{\nu}_\tau$
pumps entropy into the neutrino gas which may appear as $\Delta N_{eff}$. It is however shown in Ref. \cite{Kamada:2015era} that for $m_{Z^\prime}> 10$ MeV, the contribution to $\Delta N_{eff}$ is less than 0.1 and therefore negligible.

{\it  Effects in Supernova:}  The effects of NSI in neutrino propagation in supernova and consequences for future  supernova detection  have been studied in literature \cite{Das:2011gb}. Those results also apply to our model. Moreover $Z^\prime$ boson can also be produced inside the core and subsequently decay with $c \tau_{Z^\prime} \sim 10^{-9} {\rm km} (g^\prime /7\times 10^{-5})^{-2} (T/10~{\rm MeV}) (10~{\rm MeV}/m_{Z^\prime})^2$ which is much smaller than the supernova radius. However, its production and subsequent decay can shorten the mean free path of $\nu_\mu$ and $\nu_\tau$ and subsequently prolong the diffusion time \cite{Kamada:2015era}. In case of future supernova detection, such effects can be resolved. A more detailed study of such effects is beyond the scope of the present paper.

{\it High energy cosmic neutrinos:} The $Z^\prime$ boson can be resonantly produced via interactions of high energy cosmic neutrinos with background neutrinos which may cause a dip in the spectrum of cosmic neutrinos between 400 TeV to PeV \cite{Kamada:2015era,Cherry}. In fact, the
ICECUBE data shows hints for such a dip but further data is required to establish its presence.  From the perspective of high energy cosmic neutrinos, this model is very similar to the $\mathcal{L}_\mu-\mathcal{L}_\tau$ model studied in  \cite{Kamada:2015era}.
As shown in  \cite{Kamada:2015era}, for the parameter range of our interest ({\it i.e.,} $m_{Z^\prime}\sim 10$ MeV and $g^\prime \sim 7 \times 10^{-5}$) the optical depth can be larger than one and the dip can be therefore discerned.

{\it Muon magnetic dipole moment:} The new $U(1)^\prime$ gauge symmetry leads to a correction to the muon magnetic dipole moment similar to the one-loop $U(1)_{EM}$ correction. Up to corrections of $O(m_{Z^\prime}^2/m_\mu^2)\sim 0.01$, we can write the contribution of the $Z^\prime$ loop as $\Delta (g-2)_\mu/2= g^{\prime 2}/8\pi^2$ which for $g^\prime =7 \times 10^{-5}$ is below (but rather close to) the observational bound. This correction cannot account for the claimed discrepancy between SM prediction and measured value \cite{a-2}.

{\it Signatures at the LHC:} The components of the $H^\prime$ doublet can be produced at the LHC via electroweak interactions provided that they are not too
heavy. Their decay will produce quark jets. They can also lead to flavor changing processes at tree level. However, we have introduced a mechanism to increase $H^\prime$ mass to arbitrarily large values to avoid bounds.

{\it Neutrino scattering experiments:} We expect the effects of new physics at neutrino scattering experiments such as the NuTeV and CHARM experiments with energy momentum transfer $q$ to be suppressed by a factor of  $m_{Z^\prime}^2/|m_{Z^\prime}^2-q^2|$ relative to the SM effects. Remember that if the scale of new physics responsible for the effective potential is much larger than $q$ (as by default assumed in the literature), the correction to the cross section will be of order of main standard model. With this assumption, it was shown in \cite{bounds} that a large part of LMA-Dark parameter space is ruled out by combining the results of NuTeV and CHARM scattering experiments. These bounds do not  apply for our model because $m_{Z^\prime}^2/q^2 \ll 1$.

{\it Neutrino trident production:} The new gauge interaction can contribute to the $\mu^+\mu^-$ production in the scattering of neutrino beams off nuclei which is known as neutrino trident production:
$\nu +A \to \nu +A+\mu^++\mu^-$ where $A$ is a nucleus. The CCFR collaboration \cite{Mishra} by studying the scattering of $\sim 160$~GeV neutrino beam off an iron target and the CHARM II collaboration  \cite{trident} by studying the scattering of $\sim 20$~GeV neutrino beam off a glass target have extracted the cross section of the trident scattering process and found that the cross section is consistent with the SM prediction. From this observation, Ref. \cite{tride} has found a bound on $g^\prime$ which for $m_{Z^\prime}\sim 10 $ MeV is around $9\times 10^{-4}$. This bound is one order of magnitude above the values we are interested in.

{\it $\mu +A \to \mu+A+Z^\prime$, $Z^\prime \to \nu \bar{\nu}$:} $Z^\prime$ can be produced by scattering of muon beam off nuclei and can then decay into a $\nu \bar{\nu}$ pair. Ref. \cite{russians}
proposes using muon beam with energy of 150 GeV from CERN SPS to search for such a signal. It is shown that with $10^{12}$ incident muons, values of $g^\prime$ as small as $10^{-5}$ can be probed which tests the present scenario.
\section{Discussion and Summary}

We have presented a model leading to NSI required for the LMA-Dark solution. This model is based on a new $U(1)^\prime$ gauge symmetry. To cancel anomalies, we have suggested gauging
$\mathcal{L}_\mu+\mathcal{L}_\tau+\mathcal{B}_1-a \mathcal{B}_2-(3-a)\mathcal{B}_3$ where $\mathcal{L}_\mu$ and $\mathcal{L}_\tau$ are lepton numbers of flavor $\mu$ and $\tau$ and $\mathcal{B}_i$ is the Baryon number of flavor $i$.  The parameter $a$ is an arbitrary real number.
Taking $m_{Z^\prime}\sim 10 $ MeV and $g^\prime =(6-8)\times 10^{-5} (m_{Z^\prime}/10~{\rm MeV})$, the values of $\epsilon_{\mu\mu}^{u,d}$ and  $\epsilon_{\tau\tau}^{u,d}$ will be in the range required for the LMA-Dark solution (see Eq. (\ref{range})).
Since the $U(1)^\prime$ charges of the $\mu$ and $\tau$ flavors are the same, the mixing between $\mu$ and $\tau$ flavors respects the gauge symmetry but  mixing between $\nu_e$ and $\nu_{\mu (\tau)}$, or mixings between different generations of  quarks  break  the
$U(1)^\prime$ symmetry.

We invoke a seesaw mechanism in the neutrino sector and introduce two new scalars which are electroweak singlets but charged under the $U(1)^\prime$. Their VEVs give Majorana masses to two  right-handed neutrinos and mix $\nu_e$ with other flavors. In quark sector, we introduce a new scalar doublet $H^\prime$ with the same electroweak quantum numbers as those of the SM Higgs but charged under the $U(1)^\prime$ gauge symmetry.  The new doublet has Yukawa couplings  with quarks of different generations  and its VEV mixes them. This  new scalar doublet can contribute  to flavor violation at tree level. To avoid the lower bounds from such processes and direct searches, $H^\prime$ has to be made heavy. We have suggested a mechanism to increase the mass of this new scalar while keeping its VEV small. The mass of $Z^\prime$ boson comes from VEVs of all these new scalars (see Eq. (\ref{Zmass})).  To obtain $m_{Z^\prime} \sim 10$ MeV with $g^\prime\sim 7 \times 10^{-5}$, these VEVs (or their maximum) should be of order of few 100 GeV.

Obviously, this model,  constructed to reproduce the LMA-Dark solution,  has  effects similar to what  expected from the LMA-Dark solution on  propagation of  neutrinos in the Sun \cite{upturn} and in supernova \cite{Das:2011gb}.  Moreover, as shown in \cite{Juno}, the solution can be tested by
upcoming intermediate  baseline reactor neutrino experiments through determining the octant of $\theta_{12}$. In addition to these effects, because of  the presence of relatively light $Z^\prime$ boson
with coupling to the standard model fermions with $g^\prime >6 \times 10^{-5}$, we expect observable and testable effects in various experiments and setups: \begin{enumerate} \item  The correction to muon magnetic dipole moment is
$(g-2)_\mu/2 \sim 5 \times 10^{-11}$ which is too small to account for the claimed discrepancy. To probe such an effect, theoretical and observational uncertainties should be improved by an order of magnitude. \item The $Z^\prime$ boson can be resonantly  produced  via interactions of high energy cosmic neutrinos with the background relic neutrinos. This will create a dip in the spectrum of high energy neutrinos. The observed gap between 400 TeV to PeV might be due to such effects \cite{Kamada:2015era}. After collecting more data by ICECUBE and by other neutrino telescopes, this prediction can be tested.
 \item The $Z^\prime$ gauge boson can be produced inside the supernova core. The produced $Z^\prime$ boson will immediately decay with a decay length much shorter than the supernova core radius. This can affect diffusion time scale of $\stackrel{(-)}{\nu}_\mu$ and  $\stackrel{(-)}{\nu}_\tau$ as well as the  flavor composition and energy spectrum of supernova neutrinos.  In the event of detection of supernova neutrinos, such predictions can be tested.
\item The $Z^\prime$ boson can appear in meson decay as missing energy. The existing bounds constrain $g^\prime$ to be smaller than $10^{-3}$. To test the parameter range corresponding to  the LMA-Dark solution, precision on  the relevant processes such as [$K^+ \to (\mu^+ ~{\rm or} ~ e^+) +{\rm missing~ energy}$] should be improved by a factor of $\sim 200$.
    \item $Z^\prime$ can contribute to trident neutrino production: $\nu+A\to \nu+A+\mu^++\mu^-$. From CCFR data, there is already a bound on $g^\prime$ \cite{tride}. To probe the range of parameter relevant for LMA-Dark solution, the bound should be improved by an order of magnitude.
        \item The model can be tested by setup proposed in \cite{russians} which is based on studying $\mu+A \to \mu +A+Z^\prime$, $Z^\prime \to \nu \bar{\nu}$ with 150 GeV muon beam from SPS CERN. If the number of incident muons exceeds $10^{12}$ entire parameter range relevant for the LMA-Dark solution can be tested.
\end{enumerate}

In summary, we have presented a UV complete model which gives rise to NSI required for the LMA-Dark solution based on a new $U(1)$ gauge interaction with a gauge boson of  mass $\sim 10$~MeV. Predictions of this model can be tested
by future supernova neutrino observations, the spectrum of high energy cosmic neutrinos, trident neutrino production \cite{tride}, study of rare meson decay,
 an order of magnitude improvement in measurement and calculation of $(g-2)_\mu$ and searching for $\mu+A \to \mu+A+Z^\prime$ via the setup described and proposed in \cite{russians}.
\subsection*{Acknowledgments}
I am grateful to M. M. Sheikh-Jabbari for fruitful discussion and careful reading of the manuscript. I am also grateful to Alexei Smirnov, Michele Maltoni,  Julian Heeck, Pedro Machado and Ian Shoemaker for useful comments.
I would like to acknowledge partial support from the  European Union FP7 ITN INVISIBLES (Marie Curie Actions, PITN- GA-2011- 289442).
\


\end{document}